\documentclass{PoS}

\title{Galactic chemical evolution: \\The role of the first stars}

\ShortTitle{Galactic chemical evolution}

\author{\speaker{Gabriele Cescutti}\\
  Leibniz Institute for Astrophysics Potsdam\\
        E-mail: \email{cescutti@aip.de}}

\author{Cristina Chiappini\\
  Leibniz Institute for Astrophysics Potsdam\\
      E-mail: \email{cristina.chiappini@aip.de}}

    \abstract{The massive First Stars (the first ones to contribute to
      the chemical enrichment of the Universe due to their short
      lifetimes) are long dead, and even though efforts to directly
      observe them in high-redshift galaxies are underway, a step
      forward in this field will have to wait for JWST and ELT. The
      only way to currently validate the picture arising from the most
      modern hydro-dynamical simulations of the formation of First
      Stars is to search for their imprints left on the oldest stars
      in our Galaxy. Which imprints are we looking for? In the last
      years our group has found that many chemical anomalies observed
      in very metal-poor halo stars, as well in the oldest bulge
      globular cluster, suggest the first stellar generations to have
      been fast rotators. After giving a brief overview of the
      aforementioned results, we highlight the impact of fast
      rotating metal-poor massive stars on the chemical enrichment of
      heavy-elements such as Sr and Ba. Indeed, in fast rotating
      massive stars the s-process production is boosted. We will show,
      by means of an inhomogeneous chemical evolution model, based on stochastic approach
      to the star formation, that this fact offers a new twist in the
      interpretation of the abundance patterns and scatter observed in
      very metal-poor halo stars.}

\FullConference{XII International Symposium on Nuclei in the Cosmos\\
		 August 5-12, 2012\\
		 Cairns, Australia}

\begin{document}
\section{Introduction}

Elements with Z $>$30 are labelled neutron capture elements: they are
manly formed through neutron captures, and not through fusion, this
process beyond iron (Z=26) being endothermic.  The neutron capture
process is also split in rapid process (r-process) or slow process
(s-process) depending whether the timescale for neutron capture
$\tau_{n}$ is faster or slower compared to radioactive beta decay
$\tau_{\beta}$, according to initial definition by [1].  The neutron
fluxes so different suggest very distinct sites of production for
these elements and theoretical calculations have confirmed this idea.
The s-process takes place in the low intermediate mass stars where a
constant flux of neutrons can be provided during the asymptotic giant
branch phase; in this case, the site of production can be modelled,
and the isotopes involved are experimentally measurable; detailed
calculations are possible, even though the uncertainties, in particular
in the stellar models, can give rise to important variations in the 
resulting yields.
%see for a review, more recent calculations
The site for r-process is still under debate but all the hypotheses
lead to the common requirement of an environment in which 
extremely high neutron densities are generated.
  The general complexity of the systems, coupled with the
inaccessibility of the nuclear data of the isotopes involved, leads to
few computations with reliable yields up to now; only recently SNII
models has started to explode, so nucleosynthesis is still not
provided, with the exception of the work, on O-Mg-Ne core supernovae
(SN) by [2].

Even though the r-process is not clearly understood, all the possible
hypotheses point to massive stars and this gives rise to a different
timescale of enrichment for the two processes: few million years for
the r-process (typical lifetime for a massive star), compared to more
than half Gyr for the bulk of production of s-process elements.  It is also
expected that s-process are produced in massive stars, but this has
often been neglected due to the overall scarce amount ( in practice
negligible at very metal-poor regime) predicted by theoretical works
(see [3]). For this reason, it has been common to
associate the neutron capture elements in the extremely metal-poor
(EMP) stars of the halo with r-process production, as pointed out
already in [4].

For the first time in the work by [5], the possible importance of
s-process production in massive stars was shown in a parametrized
way. More recently, [6] used a stellar evolution code with an extended
nuclear reaction chain to confirm that fast rotating metal-poor stars
can generate heavy elements through s-process; this production of
s-process elements in massive stars provides a possible explanation of
the high abundance of Sr and Y, measured in metal-poor stars of the
Bulge, as highlighted by [7]; moreover, the concept of metal-poor fast
rotating massive stars (spinstars) has already proved to play a
crucial role in the formation of CNO elements at very low metallicity
[8].

Homogenous chemical evolution models have recently shown the
importance of spinstars for the early chemical enrichment of the
Galaxy [9,10]. However, in the case of neutron capture elements the
situation is more complex due to the presence of a huge spread in the
observed abundance ratios of the EMP stars, recently confirmed by the
results of [11].  An explanation for these inhomogeneities relies on the
stochastic formation of massive stars [12]. In this scenario, the
spread is generated by the enrichment of different species if they are
produced by different ranges of masses, providing a finite length of
the mixing zone. This has been shown for heavy neutron capture
elements vs iron in the inhomogeneous chemical evolution model by
[12]; the same approach has been used also for CNO (see [13]) to
investigate the implications of the inhomogeneous modeling in the
observed spread (in particular of the ratio N/O) in EMP stars.

It follows that also for the spread between the light and the heavy
neutron capture elements we should investigate this model and so in this
paper we use this scenario to analyze the impact of the new results
for s-process in massive stars boosted by fast rotation [6].

\section{Observational data}

We adopt observational ratios from literature; the data for the
neutron capture elements are those
collected by [14]
\footnote{http://cdsarc.u-strasbg.fr/cgi-bin/qcat?J/AN/331/474},
taking into account only the stars belonging to the Galactic halo.
Among the halo stars collected, we decide to differentiate the normal
stars from the carbon enriched metal-poor (CEMP) stars. We follow the
definition given by [15], so CEMP stars present a
[C/Fe]$>$0.9 (for the details of the categories
adopted among CEMP stars see [15]).

\section{The chemical evolution models}

The chemical evolution model is the same as adopted in [13], based on the
inhomogeneous model developed by [12].  We consider that the halo
consists of many independent regions, each with the same typical
volume, and each region does not interact with the others.  We decided
to have a typical volume with a radius of roughly 90 pc and the number
of assumed volumes is 100\footnote{We tested a larger number of
  volumes and find that our results converge after around 100 volumes
  although increasing considerably the computational time}. The
dimension of the volume is large enough to allow us to neglect the
interactions among different volumes, at least as a first
approximation.  In fact, for typical ISM densities, a supernova
remnant becomes indistinguishable from the ISM -- i.e., merges with
the ISM -- before reaching $\sim50pc$ [16], less than the size of our
region. We do not use larger volumes because we would lose the
stochasticity we are looking for; in fact, as we tested for increasing
volumes, the model tends to be homogeneous.  In each region, we
follow the chemical evolution equation and parameters as the
homogeneous model by [10].  Yields for Fe are the same as [17].  The
model takes into account the production by s-process from
low-intermediate mass stars and SNIa enrichment, as in [18].  However,
the bulk of the contribution of SNIa and AGB stars is seen mostly
above [Fe/H] $\sim -$1.5, so the impact on our results is only
marginal.

\begin{figure}
\begin{center}

\includegraphics[width=.49\textwidth]{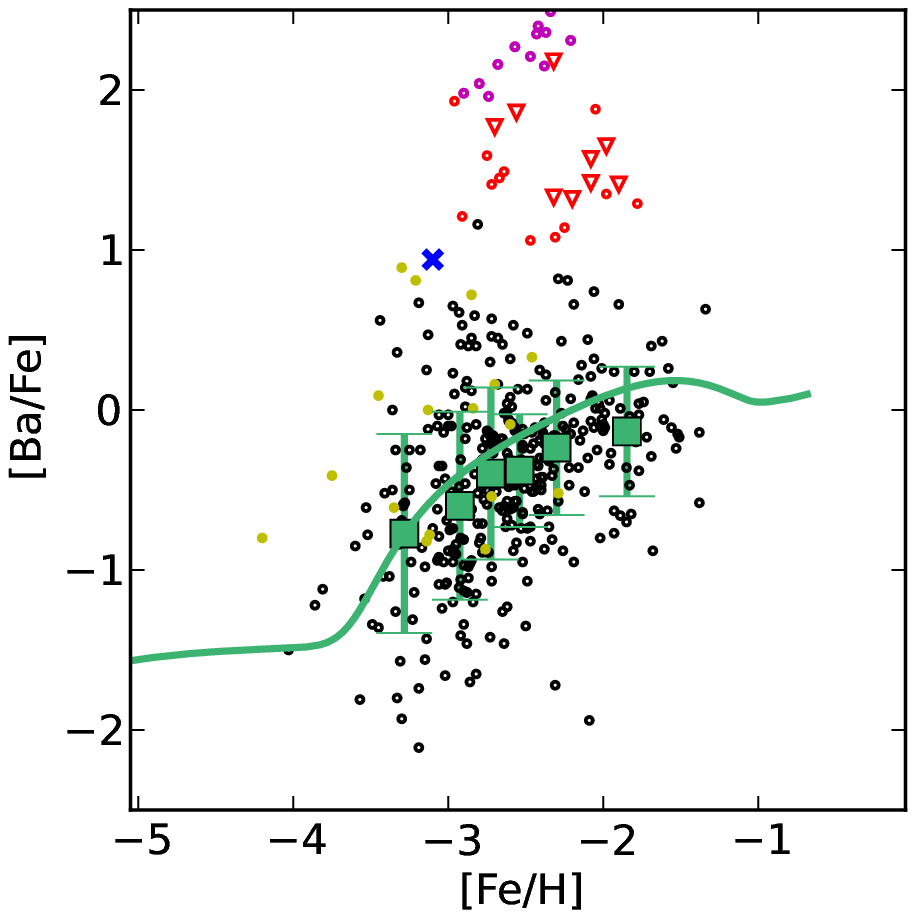}
\includegraphics[width=.79\textwidth]{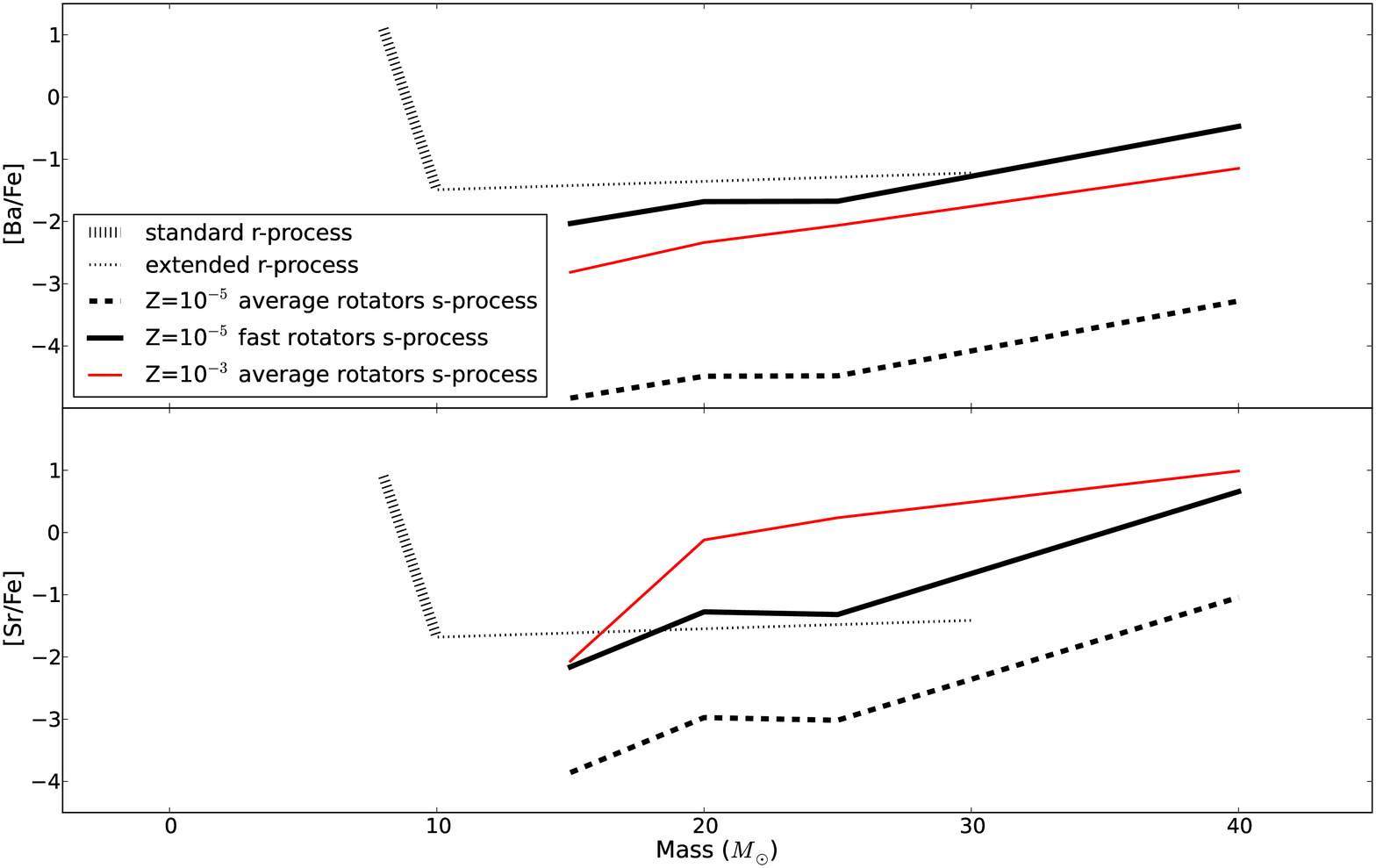}

\caption{\textit{Top:} [Ba/Fe] vs [Fe/H] abundances ratios of the
  stars gathered by [14]: black open circle are normal stars, red open
  circle for CEMP-s stars (open triangles without Eu value) magenta
  open circle for CEMP-rs, yellow filled circle for CEMP-no and blue x
  marker for the CEMP-r star.  The error bars represent the mean and
  the standard deviations for the normal stars abundances calculated
  over different bins in [Fe/H].  The bins are calculated in such a
  way each bin contain the same number of data. The results of the
  homogenous model with the assumed ``empirical'' yields for Ba is
  shown by the solid line.  \textit{Bottom:} Yields used for the
  ratios of [Ba/Fe] and [Sr/Fe], as a function of the stellar mass and
  the metallicities. The {\bf as-}, and {\bf fs-model} come from
  unpublished results by Frischknecht (2011, PhD thesis); for details
  see Sect 3.1.2.} \label{fig1}

\end{center}
\end{figure}

\subsection{Stellar yields for heavy elements}
\label{sec:yields}

\subsubsection{\emph{Empirical} yields for the r-process}

As mentioned in the Introduction, the site of production of r-process elements is still a matter of debate.
In particular, most of the neutron reactions
involving r-process elements are out of reach of nuclear physics experiment. In such a situation, we can use observational data to guide
the theoretical models for this process. 

Given the above, we adopt here the following approach (see also [18]), namely: we compute
a homogeneous chemical evolution model where the yields of Ba are chosen such as to reproduce the mean
trend of [Ba/Fe] versus [Fe/H] (see Fig.~\ref{fig1}). The latter are what we call here \emph{empirical} yields.
The \emph{empirical} yields are obtained as the simplest array able to
reproduce the observed the trend of increasing [Ba/Fe] with metallicity, but also present one important property, namely,
the need for two different sites of production (see Fig. \ref{fig1}): 
\begin{itemize}
\item a strong production in a narrow mass range 8-10$M_{\odot}$
that we call \emph{standard} r-process site;
\item an extended contribution coming from stars with larger masses (from 10 up to 30$M_{\odot}$), whose contribution
is lower by $\sim$ 2 dex than that  of the \emph{standard} r-process, and which we name here as  \emph{extended} r-process site.
\end{itemize}
Finally, to find the corresponding r-process yield for Sr, we simply scale it to Ba according to the 
solar system r-process contribution, as determined by [19].

 Interestingly the stellar mass range for \emph{standard} r-process is
 close to the one predicted by theoretical models of O-Ne-Mg
 core SN (with initial masses toward the lower end of the massive stars). Notice, however that the recent models
 [2] do not succeed in producing heavy neutron
 capture elements such as Ba.  On the other hand, similarly to the findings
 of [18], to be able to reproduce the trend of the stars with [Ba/Fe]
 $\sim -$0.7 at metallicities lower than [Fe/H] $\sim -$3, an r-process production in more massive stars is needed
in the absence of any other process able to produce such an element at very-low metallicities. As we will see, this could be different
 once the contribution of \emph{spinstars} is taken into account.

\subsubsection{The contribution of \emph{spinstars}}

To illustrate the impact of \emph{spinstars} to the chemical
enrichment of Sr and Ba in the earliest phases of the Universe, we now
focus on three sets of inhomogeneous chemical evolution models, computed
with the following set of stellar yields (see Fig.~\ref{fig1}):
\begin{itemize}

\item {\bf r-model}: assume only the \emph{empirical yields} for the
  r-process (\emph{standard} + \emph{extended}) (see Sect 3.1.1);
\item {\bf as-model}: assume only the \emph{standard} r-process yields,
  plus the s-process yields coming from rotating stellar models;
\item {\bf fs-model}: similar to the model above, but with s-process
  yields coming from fast rotating stellar (spinstars)
  models.
\end{itemize}

The nucleosynthesis adopted in the {\bf as-model} for the s-process
comes from unpublished results by Frischknecht (2011, PhD thesis).  In
this set of yields, the s-process for massive stars is computed for an
initial rotation velocity of $v_{ini}/v_{crit}$=0.4 \footnote{The
  critical velocity is reached when the gravitational acceleration is
  exactly counterbalanced by the centrifugal force} and for a standard
choice for the reaction $^{17}O(\alpha,\gamma)$; they are composed of
a grid of 4 stellar masses (15, 20, 25 and 40$M_{\odot}$) and 3
metallicities (solar metallicity,$10^{-3}$, $10^{-5}$); in
Fig.~\ref{fig1} we show the yields for the 2 lowest metallicity
cases. We do not extrapolate the production toward stars more massive
than 40$M_{\odot}$ (although it is realistic to have a production also
in this range), but we extended the $Z=10^{-5}$ grid down to Z=0.  In
addition to the s-process in massive stars, we take into account our
empirical r-process enrichment but coming only from the
\emph{standard} r-process site. In this way we are decoupling the
sites of production for the two processes. Interestingly, this figure
suggests that the role of the ``extended r-process'' site can be
played by the s-process of \emph{spinstars}.

It is worth to point out that the yields we have used for the {\bf as-model}
are very conservative among the models computed by [6] for
25$M_{\odot}$.  A more extreme case of s-process production is what we
used in the {\bf fs-model}. This case is achievable if one considers the
s-process yields for massive stars computed for a rotation rate of
$v_{ini}/v_{crit}$=0.5 and for a reaction rate $^{17}O(\alpha,\gamma)$
one tenth of the standard choice.  We do not have a fully computed
grid for these parameters, but we have scaled the previous yields
guided by the results obtained with these parameters for a
25$M_{\odot}$ of $Z=10^{-5}$ by [6] (cfr. in their paper Table 2), and
applying the same factor to all the masses at $Z=10^{-5}$, as shown in
Fig. \ref{fig1}. Again, for this {\bf fs-model},
in addition to the s-process in \emph{spinstars}, we take into account
the contribution by our empirical \emph{standard} r-process.

\begin{figure*}[ht!]
\begin{minipage}{150mm}
\begin{center}
\includegraphics[width=150mm]{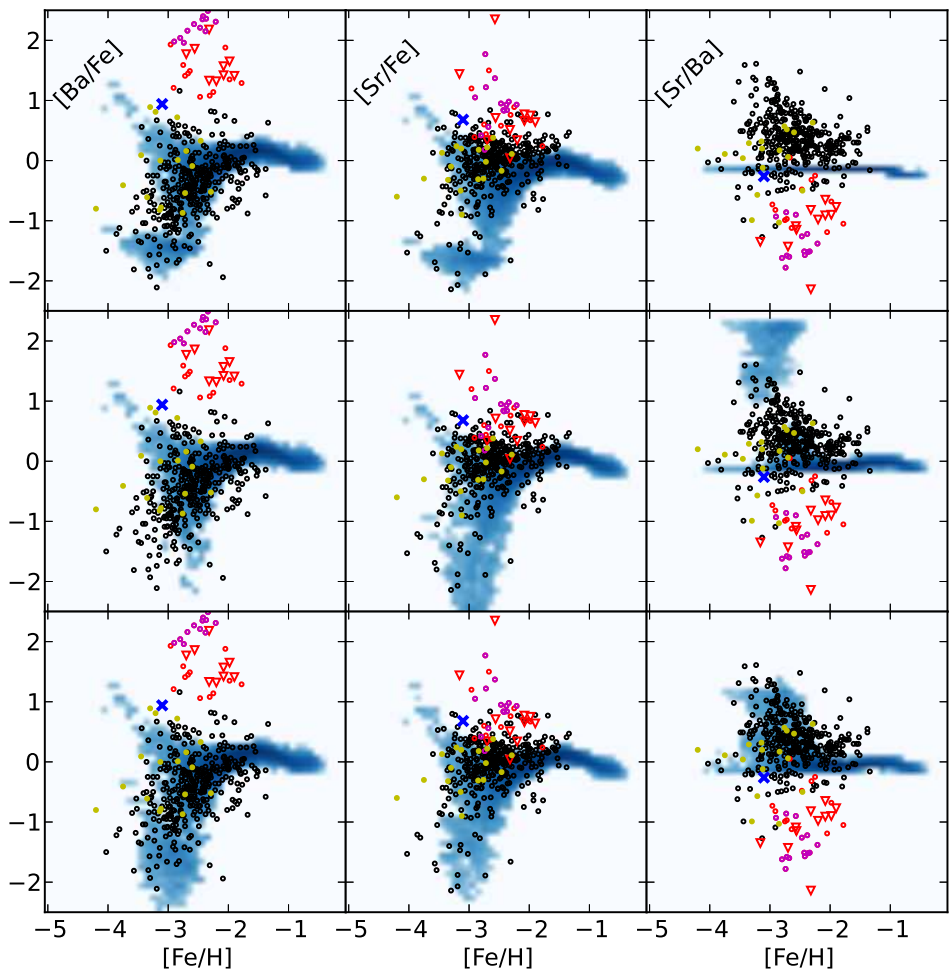}

\caption{On the left [Ba/Fe], on the center [Sr/Fe], on the right [Sr/Ba],
 vs [Fe/H]. Upper, center and lower panels for {\bf r-}, {\bf as-} and {\bf fs-model} respectively.
The density plot is the distribution of simulated  long living stars for the model.
Superimposed, we show the abundances ratios of the stars
gathered by [14]. The symbols adopted are the same as Fig.~1 (top panel).
}\label{fig2}

\end{center}
\end{minipage}
\end{figure*}

\section{Results}
%\subsection{r-model}

Our main results are summarized in Fig.\ref{fig2}.  We start by analyzing the results of the {\bf r-model}
(upper panel), which assumes only the contribution from massive stars
via our \emph{empirical} r-process yields.  The spread obtained by this
model matches the dispersion of [Ba/Fe] (upper, left panel), confirming that the hypothesis of a contribution
by two distinct sites seems to be a good one to explain the data. In the particular case of this model, the two-sites
of production are illustrated by a
\emph{standard} r-process which takes place in the lower mass range of
the massive stars, and an \emph{extended} r-process taking place in
more massive stars. The first one is assumed to be much more efficient
(larger quantities of ejected material) than the second.

A good match of the observed [Sr/Fe] is also obtained by the same
model, especially at very metal-poor metallicities.  As the Sr yields
adopted here are obtained just by using the Ba/Sr ratio matching the
observed solar system r-process [19], this ratio is simply
constant with metallicity (as shown in the upper, right panel). This
suggests that the some other physical process, taking place in the
same mass range of what we have called \emph{extended} r-process might
be contributing to produce part of the Sr and Ba in the very early
Universe (as the scatter is larger between metallicities $-$2.5 and
$-$3.5). In other words, the {\bf r-model} is useful to highlight the
issue we are trying to solve in the present work.

Lets now turn to the results obtained with our {\bf as-model} (see
Fig \ref{fig2}, second row), and see the impact of the
s-process production of Ba and Sr by rotating massive stars. In
this case we added a rather conservative s-process production from
\emph{spinstars}, and turned off the contribution of what we have
called \emph{extended} r-process.  The results for [Ba/Fe] and [Sr/Fe]
are not completely satisfactory as the model cannot reproduce the low
[Ba/Fe] ratios observed in extremely metal-poor stars with [Fe/H]$<-$3
(left and middle panels of the second row). This happens because this
rather conservative model predicts not enough amounts of Sr, and
particularly, of Ba (this can be seen in Fig.~\ref{fig1} by comparing
the stellar yields of the \emph{extended} r-process -- dotted line --
with the yields of the {\bf as-model} at $Z=10^{-5}$ -- dashed line).

Despite the shortcomings described above, the interesting result of
this model resides in the [Sr/Ba] plot (second row, right panel),
where it is clear that this new process produces an overabundance and
creates a spread in [Sr/Ba] at [Fe/H] $\sim -$3.  From this one can
conclude that the s-process produced in rotating massive stars
seems to act in the correct direction.  Nevertheless, these results
are affected by the ratio of the yields of Sr and Ba predicted by this
specific stellar evolution model (similar to model B1 of [6]). These
prescriptions produce [Sr/Ba]$>$2, whereas the EMP stars show at
maximum [Sr/Ba]$\sim$1.5.

Finally for the {\bf fs-model} (Fig \ref{fig2}, bottom panels), where
we adopted less conservative stellar models predicting even larger
s-process enrichment (essentially due to a larger $v_{ini}/v_{crit}$
and a lower $^{17}O(\alpha,\gamma)$ reaction rate similar to the model
B4 of [6] -- see Section~\ref{sec:yields}), the
theoretical predictions and observations show an striking agreement.
This model not only reproduces well the [Ba/Fe] and [Sr/Fe] scatter
(left and middle bottom panels), but can also account for the observed
[Sr/Ba] spread (right bottom panel), at the right metallicity
interval.
This shows that with a less conservative production of s-process in
fast rotating massive stars (as is the case in the {\bf fs-model}),
this process could play the same role as our \emph{extended} r-process
in the {\bf r-model}. In addition, it shows that a \emph{standard}
r-process (taking place in the lower mass range of the massive stars)
enters at play with a weight that increases as the metallicity
increases (still in the very metal-poor range).

\begin{figure}[h!]
\begin{center}

\includegraphics[width=.79\textwidth]{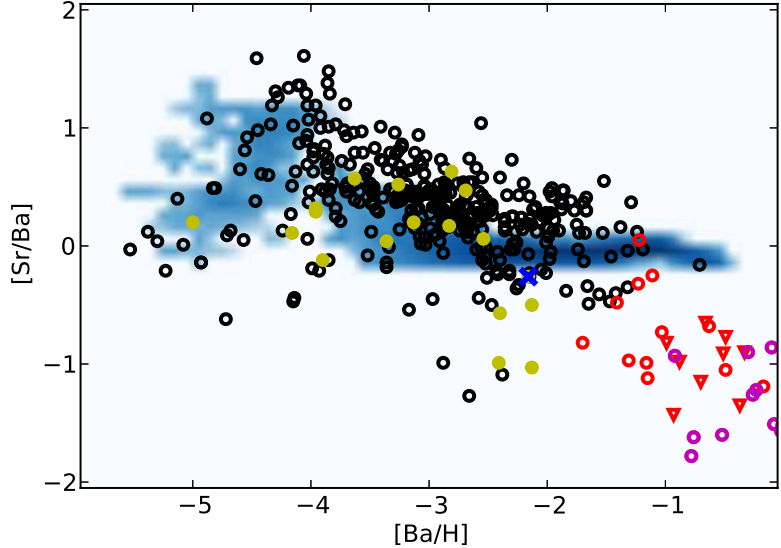}

\caption{[Sr/Ba] vs [Ba/H], the density plot is 
the distribution of simulated  long living stars for {\bf fs-model},
Superimposed, we show the abundances ratios of the stars
gathered by [14]. The symbols adopted are the same as Fig.~1.
}\label{fig3}

\end{center}
\end{figure}

In Fig. \ref{fig3}, we show the same results presented in the lower
right plot of Fig.\ref{fig2} but using [Ba/H] on the x-axis. Indeed
[Fe/H] on the x-axis is the most common way to present these figures
but using instead the [Ba/H] we avoid the assumption regarding the
iron yields. The point is that these yields are naturally bounded to
the still not completely understood SNII explosion mechanism.  Our
model reproduces also in this case the peculiar behavior seen in EMP
stars, namely, a high ratio of [Sr/Ba] at low [Ba/H], and the correct
amount of scatter, proving that the assumption on iron yields is not
influencing our results.

\section{Discussions and conclusions}

We have developed an inhomogeneous chemical evolution model for the
halo with the aim of explaining the observed scatter (or lack of) in
the abundance ratios of key chemical elements in very metal-poor
stars. The models presented here serve as a test-bench to study the
different nucleosynthetic prescriptions proposed by different stellar
evolution groups.  The goal is to identify key abundance ratios, as
well as the most promising stellar yields, to be then implemented in
cosmological simulations (a project that we are already pursuing).  

We first show that one can predict the observed spread of
  [Ba/Fe] and [Sr/Fe]
in EMP stars with an inhomogeneous chemical
  evolution model taking into account the stochastic formation of
  stars of different masses in the early phase of the Galactic
  halo. The latter assumption, when coupled with a production of
  neutron capture elements coming from: a) a relatively rare, but
  efficient, site of production (here illustrated by a contribution of
  stars in the 8-10 M$_{\odot}$ mass range), and b) a second site of
  production, which is less efficient producing lower amounts of
  neutron capture (here illustrate with a contribution from stars in
  the 10-40 M$_{\odot}$ mass range). More importantly, we show that
  despite the good agreement with the observed spread of [Ba/Fe]
  and [Sr/Fe], an extra process is needed to explain the observed
  [Sr/Ba] scatter.

%\item 
The presence of r-process rich stars, with a common strong
  r-process signature, is an observational constraint for the first
  site of production: if this site produces strong enhancement in a
  relatively rare events, then we can provide a solution for these
  stars. Here we show that keeping fixed this first site of
  production, but adding the contribution to the neutron capture
  elements by fast rotating massive stars (now considered as second
  site of production), then it is possible to create a scatter in the
  ratio of [Sr/Ba], as observed.

  In particular, if we consider stellar models for fast rotating
  massive stars ({\bf fs-model}) that are less conservative in their assumptions, we
  are then able, for the first time in our knowledge, to reproduce
  simultaneous the [$\alpha$/Fe], [Sr/Fe], [Ba/Fe] and [Sr/Ba] trends
  and scatter observed in halo stars with an inhomogeneous chemical
  evolution model.  Although the proposition that the scatter in the
  observed spread between heavy and light neutron capture elements in
  EMP stars could be due to the contribution of \emph{spinstars} had
  been made before ([7]), here we show for the first time quantitative
  estimates that seem to confirm this hypothesis. Notice that in this
  elegant solution, we are able to explain the results with just two
  different sites of production for Sr and Ba, without requiring more
  complicated scenarios.

Our results also show that to reconcile the model to the
  observations we need an r-process site of production decoupled from
  the one of s-process. We assumed here the most straightforward
  solution, which is to consider two different mass ranges for the two
  processes, but other solutions providing a large amount of r-process
  production in only a fraction (roughly 15\%) of the massive stars
  would be equally valid, as in the case of magnetorotationally driven
  supernovae, described by Thielemann in his contribution.

\acknowledgments
We thank Urs Frischknecht, Raphael Hirschi and George Meynet for
having provided the nucleosyntesis yields in advance of publication.

\vspace{0.75cm}

\noindent {\bf Questions}

\bigskip

\noindent \textit{Chiaki Kobayashi}: Can you explain the details of
your inhomogeneous models because the results, in particular x-axis,
highly depend on these?  

\smallskip

\noindent \textit{Gabriele Cescutti}: We set the parameters of our
chemical evolution model to be able to reproduce the MDF observed in
the halo. The fact that the model is formed by several independent
volumes in which the stars are chosen stochastically (cumulatively
following the assumed IMF), does not change the overall trend and the
average over the different volumes is still reproducing the MDF (see
Fig.2). Moreover, if we use in x axis the [Ba/H] relaxing the issue
in the assumption of the iron yields, we are still reproducing
the observational data.

\smallskip

\noindent  \textit{Chiaki Kobayashi}: Can you also explain the
distribution of [C/Fe] and [N/Fe] as well? 

\smallskip

\noindent \textit{Gabriele Cescutti}: Using the same chemical
evolution model, we have shown in Cescutti \& Chiappini 2010, 
that the spinstars predictions for CNO are able to reproduce
 the [C/Fe] and [N/Fe].

\smallskip

\noindent  \textit{Brad Gibson}:
What is your predicted distribution of even-to-odd Ba isotopes
in the halo? How does it compare to what (little) we know about such
ratios (eg. Mashonkina et al 2003)?

\smallskip

\noindent \textit{Gabriele Cescutti}: For the stars showing a high
[Sr/Ba] ratio at [Fe/H]$< -$2 we expect to find a isotopic fraction
typical of the s-process, being their abundances produced by
s-process from spinstars, according to the presented model.  The
comparison with the results of Mashonkina et al. (2003) is not
straightforward, since they measured isotopic ratios for stars of
higher metallicity. Nevertheless, our model predicts in that
metallicity range a mixture of r- and s-process,
which is in agreement with the results by Mashonkina et al. (2003).

\end{document}